\documentclass[12pt,preprint]{aastex}

\shorttitle{Disk-Jet Interaction Model in Microquasars}
\shortauthors{L.Nobili}

\newcommand{\eps}{\varepsilon}
\newcommand{\rin}{\hbox{$\, r_{in}\, $}}

\newcommand{\nk}{\nu_k}
\newcommand{\thetao}{\theta_{obs}}

\begin{document}

\title{A Disk--Jet interaction model for the X--Ray Variability in Microquasars}
\author{L. Nobili}
\affil{Department of Physics, University of Padova, \\
Via Marzolo 8, I--35131 Padova, Italy \\
e--mail: nobili@pd.infn.it}

\begin{abstract}
We propose a simple dynamical model that may account for the
observed spectral and temporal properties of GRS 1915+105 and  XTE J1550-5634. The model is based on the
assumption that a fraction of the radiation emitted by a hot spot lying on the accreting disk
is dynamically Comptonized by the relativistic jet that typically accompanies the microquasar
phenomenon. We show that scattering by the jet produces a detectable modulation of the
observed flux. In particular, we found that the phase lag between hard and soft photons
depends on the radial position of the hot spot and, if the angle between the jet and the line of sight is
sufficiently large, the lags of the fundamental and its harmonics may be either positive or negative.
\end{abstract}


\section{Introduction}\label{sec-intro}

Microquasars are a subset of the broader class of X--ray novae in which the flaring radio counterpart
is resolved into relativistic jets.
It is universally acknowledged that they are binary galactic systems formed by a
black hole (BH) accreting mass at the expense of a donor-star. Remarkable examples of these
systems are given by GRS 1915+105 \citep{gre96,mor97,che97,mun99,bel00} and XTE J1550-564
\citep{wij99,sob00}.
Their distinctive features are the extremely wide range
of variability modes and the presence of alternating phases of dramatic variability and of
remarkably regular behavior. Quasi Periodic Oscillations (QPOs) have
been observed in an incredibly large interval of frequencies.
In the case of GRS1915+105, radio observations show superluminal
expansion \citep{mir94}, usually interpreted in terms of emission from gas expanding at relativistic
bulk speed \citep{fen99}. The more recent mass estimate of the BH gives
$\, M\approx 14 \, M_\odot$ \citep{gre01}.

The X--ray flux consists of a sum of a soft thermal component from the accreting disk (THC),
peaking at a temperature $\,\lesssim 1\, $ keV, and a high energy component (HEC) extending
up to $\,\approx 100\, $ keV.
This last part of the spectrum has often, but not always, the form of a power--law tail.
The relative amplitude of these two components is strongly variable and this
can be explained in terms of a sudden destruction
and subsequent slow replenishment of the inner part of the accreting disk \citep{bel97}.
This event marks the transition from a state which is characterized in large part by a
multi--temperature black body emission from a standard disk,
to a state in which the emission is due to a semi-evacuated disk lacking of its most luminous inner region.

While it is generally accepted that the THC is associated to the disk, the origin of the hard tail is
presently less certain. In fact, although the upscattering of photons by a hot corona can
account for many of the observed properties, 
there is still an open debate about the possible relevance of the Comptonization by bulk motion.
\citep{zdz01,tit02,rei01}, or of repeated Compton scattering of seed
photons from the disc off electrons with a hybrid, thermal/non-thermal distribution \citep{gie99,zdz01}.
Both processes, in fact, produce similar effects, and any attempt to fit observational
data with theoretical results rarely gives unambiguous answers. Moreover, all current models suffer from
a number of drastic approximations and contain so many geometrical and physical parameters (mass of the
BH, accretion rate, size of the disk, location, density and temperature of the corona, etc.) to make
extremely difficult any accurate quantitative comparison.

A possible solution of this distressing situation can be found by considering second order effects,
and among these we count time lag measurements.
In fact, an astonishing feature of the proto-typical microquasar GRS 1915+105 is the tight correlation between
the 0.5--8 Hz QPO and its spectral properties  and variability\citep{rei00}.
Even more impressive is the gradual decrease of time lag between hard (5-13 keV) and soft (2-5 keV)
photons from initially positive values (hard photons arrive later than soft photons)
to negative values (soft photons are delayed).
It is well known that thermal Comptonization by a uniform
hot corona cannot account for negative lags as well as for the observed decrease in the hard
color as the disk inner radius \rin\ moves inwards. To circumvent this problem \citet{nob00} proposed
that the evaporation of the inner portion of the accreting disk leads to the formation of an
approximatively spherical, non-uniform cloud filling the central region.
During the replenishment of the semi--evacuated part of the disk the cloud, assumed not uniform
in temperature and density, becomes more and more compact, following the inward motion of the disk edge.
Initially the cloud is large and marginally thick to electron-scattering. Photons generated by the
disk are upscattered in the inner and hotter regions, emerging with a positive time lag. In later stages,
the cloud is much more dense, with the consequence that in its central regions photons reach a
Bose-Einstein equilibrium, and are then downscattered when they propagate through the outer and cooler shells.
In this latter configuration lags turn out to be negative.
Their calculations succeeded to reproduce some of the features of GRS 1915+105,
but cannot explain, e.g., the apparent alternating negative and positive phase lags between the
harmonics of the QPOs observed in some microquasars \citep{wij99}.

In this paper we propose a simple alternative model, showing that also dynamical Comptonization
by the jet on the radiation emitted by an accretion disk can account for some of the observed
properties of the source. The model is alternative also
to other interpretations of the timing properties of microquasars based on dynamical Comptonization
effects in converging flows \citep{lau01}.
The geometrical setting of our model is discussed in the next section,
while in section 3 we present a numerical simulation based on a Monte Carlo technique.

\section{The Model}\label{the-model}

\begin{figure}
\plotone{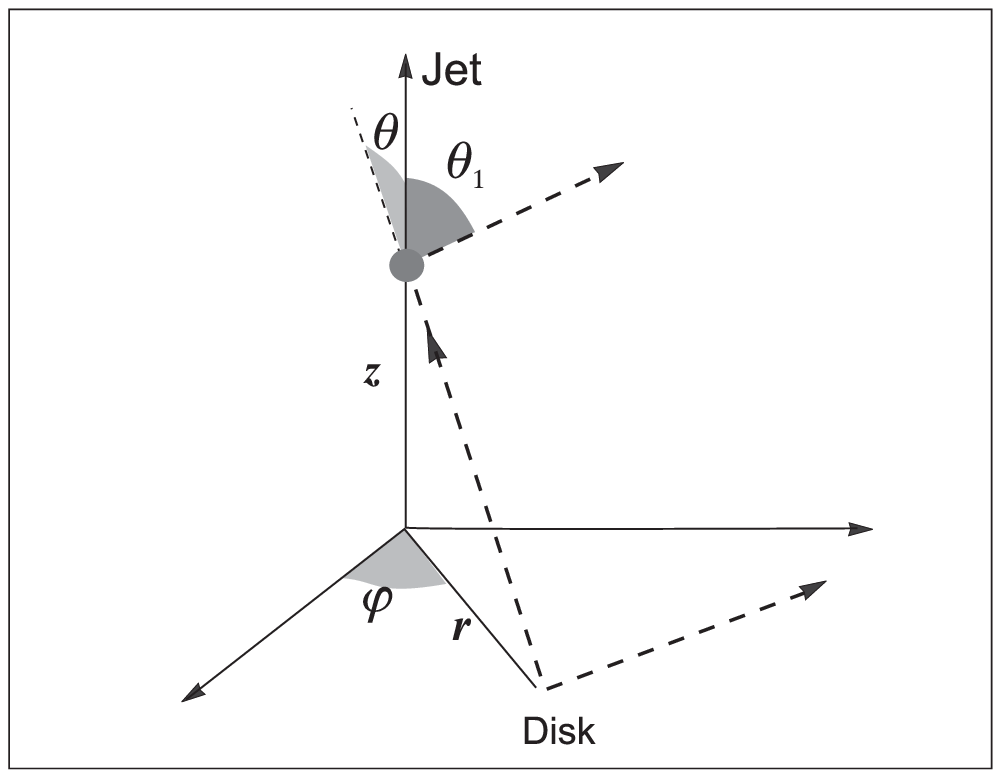}
\caption{A schematic view of the model discussed in the text.\label{Fig1}}
\end{figure}

One of the defining characteristics of microquasars is the fact that they exhibit relativistic ejection
of matter. From the observations of superluminal expansion in GRS 1915+105,  \citet{fen99} inferred a
bulk ejecta velocity $\,\beta=v/c \gtrsim 0.95\, $ and a mass outflow rate
$\dot m  \gtrsim 10^{18}\, {\rm g\, s^{-1}}\approx {\dot M}_{Edd} $.
The question of what the accelerating mechanism really is, is still unanswered. The most probable
hypothesis is that the plasma ejection is driven and collimated by magnetic fields anchored to the disk
(see, e.g., \citet[]{hei00}).
This implies that the accelerating engine should be effective near the central black hole, thus making
reasonable the assumption that the plasmoids do attain their maximum velocity already at a few gravitational
radii. A rough calculation shows that the inner portion of a jet with an opening angle
$\Theta_J\approx 0.2 $ radians becomes optically thick to electron scattering for
$\,\dot m \gtrsim 0.1\, {\dot M}_{Edd}$, 
we then argue that during the active state
a small but non-negligible fraction of the photons emitted by the disk is scattered by the
electrons on the jet. It should be noted that, whether a photon acquires
or loses energy depends on the angles of scattering and ultimately, on the position of the source, the height
where scattering occurs and the inclination with respect to the observer.
As an example, let us consider a monochromatic source of photons lying on the equatorial plane at a radial
distance $\, r$. For reasons of simplicity, let us assume that photons directed toward the axis undergo a
single scattering at some height $z$ before leaving the region, as illustrated schematically in
Figure~\ref{Fig1}. If $\eps= h\nu\, $ and $\eps_1=h\nu_1\,$ are the initial and final photon energy respectively,
and $\xi=z/r$, we have
\begin{equation}\label{compt}
\eps_1=\frac{1-\beta\cos\theta}{1-\beta\cos\theta_1}\eps = \frac{\eps}{1-\beta\cos\theta_1}
\left( 1-\frac{\beta\,\xi }{\sqrt{1+\xi^2}} \right)
\end{equation}
where $\,\theta$ is the angle between the axis and the photon direction before the scattering, and
$\,\theta_1$ is the fixed angle of observation. From equation~(\ref{compt}) we see that the ratio $\eps_1/\eps$
may be either larger or smaller than unity, depending on the value of $\,\xi$. Then photons may gain or lose
energy depending on the height at which they are scattered.
Of course, in a less schematic view, we need to consider the emission from the whole disk. Moreover
scattering does involve an extended portion of the jet and obeys to a probabilistic law that depends on
the gas density. As it will be shown in Section~\ref{sec-monte}, adopting a more realistic
configuration and assuming a cold relativistic jet (i.e. the electron thermal velocity much lower than the
bulk velocity) Comptonization by bulk motion leads to the formation of a diluted black--body
energy distribution. This distribution is roughly similar to the familiar multicolor spectrum of the
unperturbed disk, but differs from it because its maximum is shifted to higher frequencies
and it has a more extended hard tail.
One cannot exclude, however, the possibility that the inner portion of the jet is heated by, e.g,
turbulence, radiative friction or magnetic dissipation. If these mechanisms are capable of maintaining
electrons in nearly virial equilibrium with the bulk motion, the combination of thermal and dynamical
Comptonization leads to the formation of a power--law tail that is practically indistinguishable from that
produced by a hot corona.

This model has also implications on the timing properties of the observed radiation.
In fact, photons scattered at different $\,z$-coordinates describe different path lengths, arriving
at infinity with different energies and delays. A straightforward calculation yields the following
functional relation between the phase delay 
(in radians) and the geometrical parameters
(we ignore an inessential constant related to the travel time of the radiation from the source to the Earth):
\begin{equation}\label{time}
2\pi\nk (t_{obs}-t)
= \frac{1}{\sqrt{r}}\left[\sqrt{1+\xi^2}-\xi\cos\theta_1+\sin\theta_1
\,\sin\phi\right] \end{equation}
where $\phi$ is the angular coordinate of the point source, $t\, $ and $\, t_{obs}\, $  the time
of emission and observation, respectively.
In equation~(~\ref{time}), $m=M/M_\odot$ and  $r$ is in units of
$r_g=GM/c^2$  (here and in the following all distances will be expressed in gravitational units).

\begin{figure}
\plotone{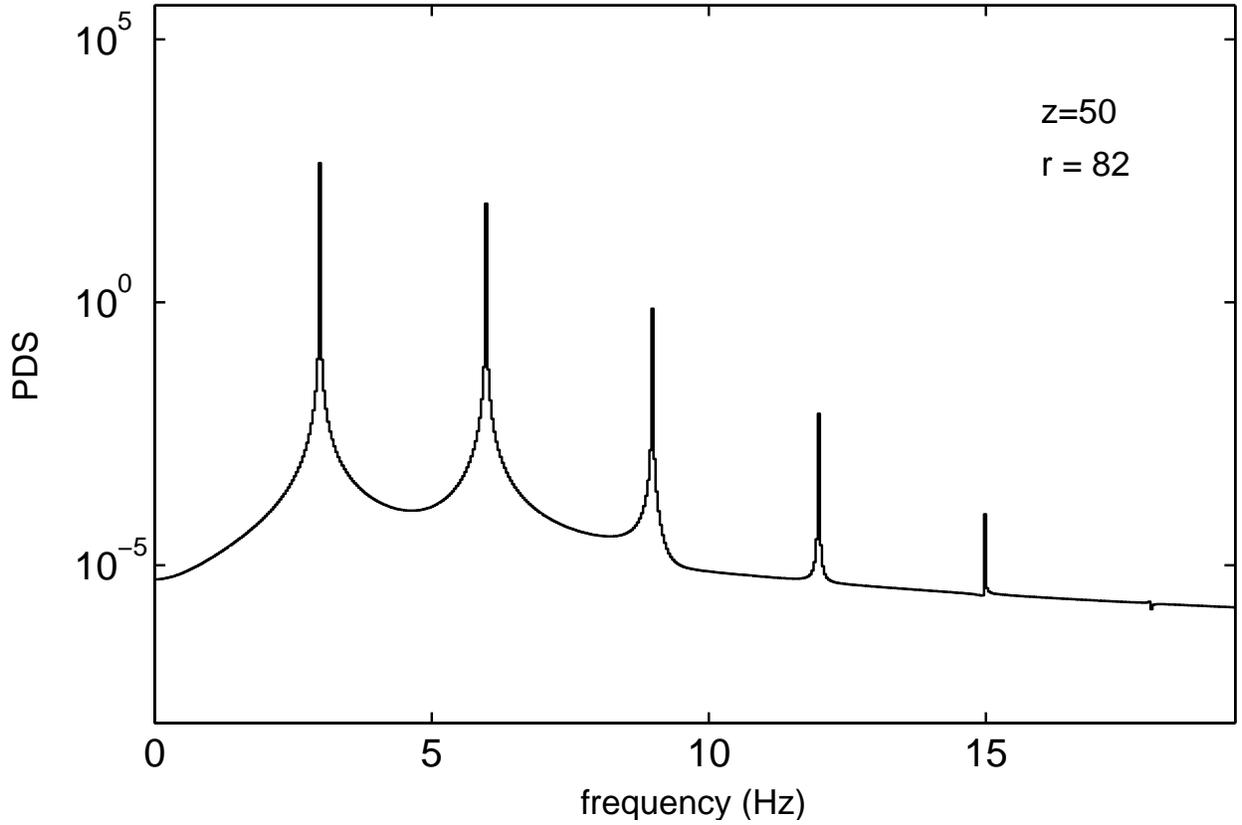}
\caption{Power density spectrum of the function $1+\cos^2\Theta'(t(t_{obs}))\, $ which describes
the scattered component of the observed flux [cfr.eqn.(\ref{Theta})].
The strong non-linear dependence of this function on $t_{obs}$ leads to the simultaneous formation of a forest
of sub-harmonics (see text for details).\label{Fig2}}
\end{figure}

\begin{figure}
\plotone{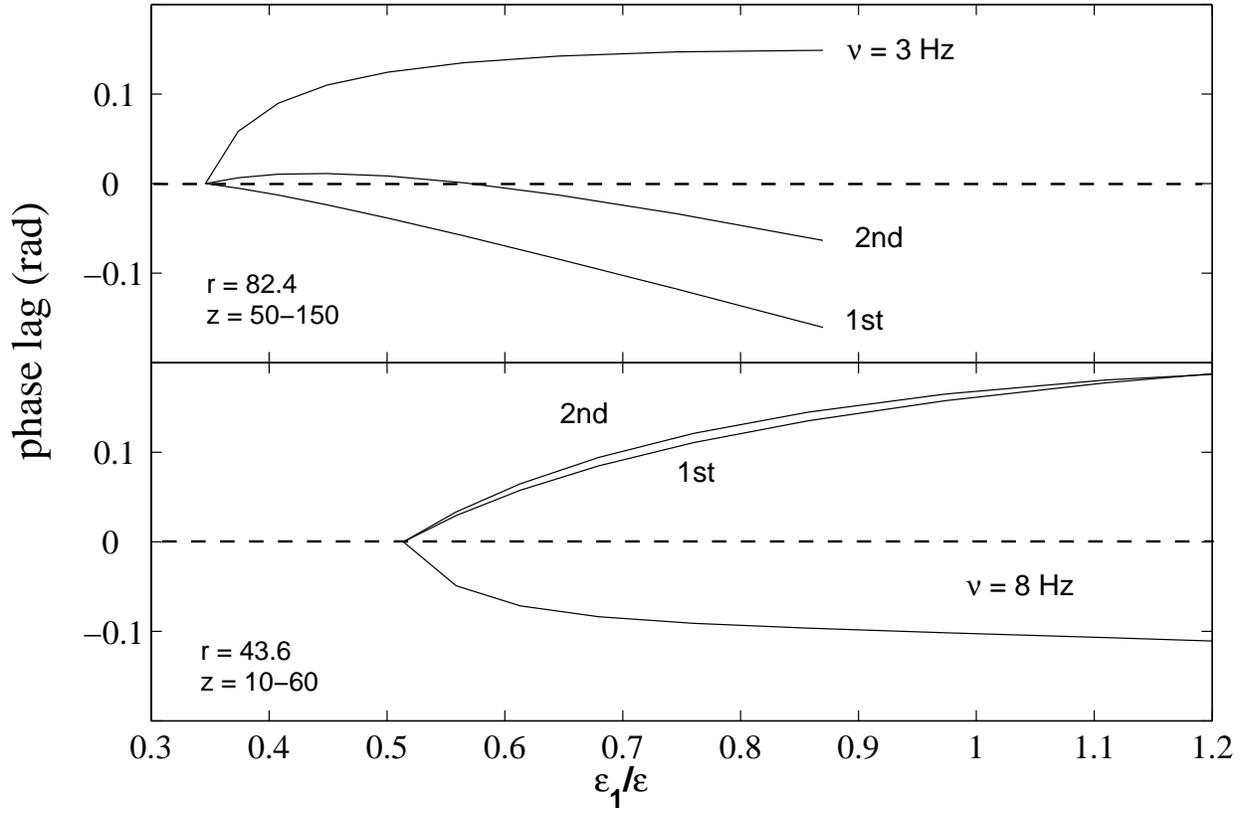}
\caption{Phase lag vs. energy for the fundamental and the first two harmonics
for two different distances $r$ of the source from the axis.
Phase lags are normalized to those of the lowest energy. Both $r$ and $z$ are in units of $r_g$.
\label{Fig3}}
\end{figure}

Let us consider now an orbiting source of radiation localized on the disk, such as a hot spot,
a spiral shock, or any other
steady perturbation, lasting with an approximate constant flux for a time much longer than the orbital
period $\nk^{-1}$, with $\nk = 3.22\times 10^4 m^{-1}\, r^{-3/2}$ Hz.
The formation of this kind of instabilities, where a substantial fraction of the accretion energy is
dissipated, is discussed e.g. by \cite{tag99}.
They are associated with the development of powerful standing waves in
moderately magnetized disks and are similar, in some aspects, to the Great Red Spot in Jupiter.
Because the differential (Thomson) cross section is proportional to $1+\cos^2\Theta'$, with
\begin{equation}\label{Theta}
\cos\Theta'(t) = \cos\theta'\cos\theta'_1 -\sin\theta'\sin\theta'_1 \sin(2\pi\nk t),
\end{equation}
the total flux measured by distant observer consists of the superposition of a nearly constant background
plus a modulated component due to the reprocessed radiation. In equation~(~\ref{Theta}) primed angles are
measured in the electron local rest frame, and are related to the corresponding unprimed quantities via
the standard Lorentz transformations. For the sake of simplicity, and because we are mainly
interested in the non linear effects in the harmonic functions, we will not consider here the possibly weak
variation of luminosity associated with the orbital revolution of the source and caused, for instance, by
the kinematic Doppler shift. We point out that, in addition to the quadratic dependence
given by equation~(~\ref{Theta}), there is a further element of non linearity in the signal. This effect is caused
by the dependence of the arrival time of the signal on the angular coordinate $\,\phi$ of the source.
At any given instant $\, t_{obs}$, the {\it scattered\/} component of the observed flux varies as
$\, 1+\cos^2\Theta'(t(t_{obs}))$, where  $\, t(t_{obs})\, $ is a non linear function of the time of
observation, obtained by inverting equation~(~\ref{time}) with $\phi=2\pi\nk t$. As shown in
Figures~\ref{Fig2} and \ref{Fig3}, this
double non linear variation generates a number of harmonics with different signs for the lags.
The curves were obtained simply applying a standard PDS technique to the function
$\,\cos^2\Theta'(t_{obs})$.
These figures are only representative, because the strength and the relative sign for the lags
of the fundamental and sub-harmonics turn out to depend sensibly on the model parameters, i.e.
the outflow velocity $\beta$, the place of emission $r$, the range in $z$ where scattering occurs and,
finally, the direction $\theta_1$ between the mean electron bulk velocity and the line of sight.
Although the first and second harmonics seem to exhibit preferentially lags of the same sense,
we cannot exclude that a different parametrization could produce lags with opposite sign.

\notetoeditor{Figures 1 and 2 should appear side-by-side in print.}

Unfortunately the sensitivity of the results to so many parameters is rather troublesome and represent a
severe obstacle for a reliable test of the model. Yet, we cannot exclude that precisely these intricate
relationships cause the complex evolutionary behavior of microquasars.

\section{A Monte Carlo simulation}\label{sec-monte}

The results obtained in the previous section can qualitatively explain some spectral and temporal
properties of microquasars. 
In particular, Figure~\ref{Fig3} seems to suggest the existence
of a time correlation between different harmonics, but no definite conclusion can be drawn,
for example, about the absolute value of the delays.
In fact, a number of geometrical and physical effects have been neglected in the
previous analysis and need to be included in a more quantitative study.
Among others, we mention:
1) a non monochromatic spectrum from the point source;
2) a thermal component of the electron velocity in addition to the bulk velocity;
3) an extended portion of the jet where radiation is effectively scattered;
4) the dependence of the (multi) scattering processes on the optical depth, $\tau_{\ell}$, evaluated
along the effective optical path.
Moreover, it is likely that the formation of a hot corona above the disk modifies the spectral
distribution of photons causing, at the same time, an additional (positive) time delay. Non uniform
electron velocity, density and temperature distributions along the jet are additional causes of uncertainty.

To carry out a more quantitative, though preliminary test, we used a
Monte Carlo code to study the radiation emitted by a Shakura-Sunayev disk \citep{sha73},
without any color correction, plus a conical jet
expanding with a constant bulk velocity $\,\beta=v/c=0.95$ for $\, z>z_{min}=15$.
The exact value of $z_{min}$ turns out not to be critical unless the inner edge of the disk is near the
BH horizon. 
The disk was truncated at $r_{in} \geq 6$. The code is fully general-relativistic, i.e. photon trajectories
are evaluated following the effective null geodesics in a Schwarzschild metric, with a mass of the
central BH $\, M=14 M_\odot$.
Compton scattering is calculated using the Klein-Nishina cross section and taking into account
for both thermal and electron bulk motion. For $\, z>z_{min}\, $ the gas density is assumed to vary
with the polar coordinate $\,\theta_p\,$ as $\,\rho \propto \rho_0\cos^{n}\theta_p/z^2$, with $\, n=10$.
This strong angular dependence was introduced for computational reasons and for ensuring an appropriate
narrow configuration of the outflow. As expected, the results of our computations depend sensibly on the
optical thickness $\tau_\ell\propto\rho_0$ (or the mass loss rate $\,\dot m$), but are only weakly
dependent on $n$.

In this preliminary investigation our main goal is to evaluate the apparent time lag of the
signal between soft and hard bands and the radiation spectrum generated by the whole
configuration (i.e. disk plus jet). The best results were obtained considering the outflow to be hot and
adiabatically expanding. In particular, we have adopted a temperature at the base of the jet
$T_{0}\approx 30$ keV, decreasing outwards as $\, r^{-4/3}$.
This implies that far from the central black--hole only Comptonization by bulk motion is effective.
Higher values
of the temperature or higher Lorentz factor are clearly admissible and would cause the formation of a harder
or even inverted tail in the spectrum.
However, because an accurate comparison of our results with observations is out of the scope of the present
investigation, we have not varied this parameter.

\begin{figure}
\plotone{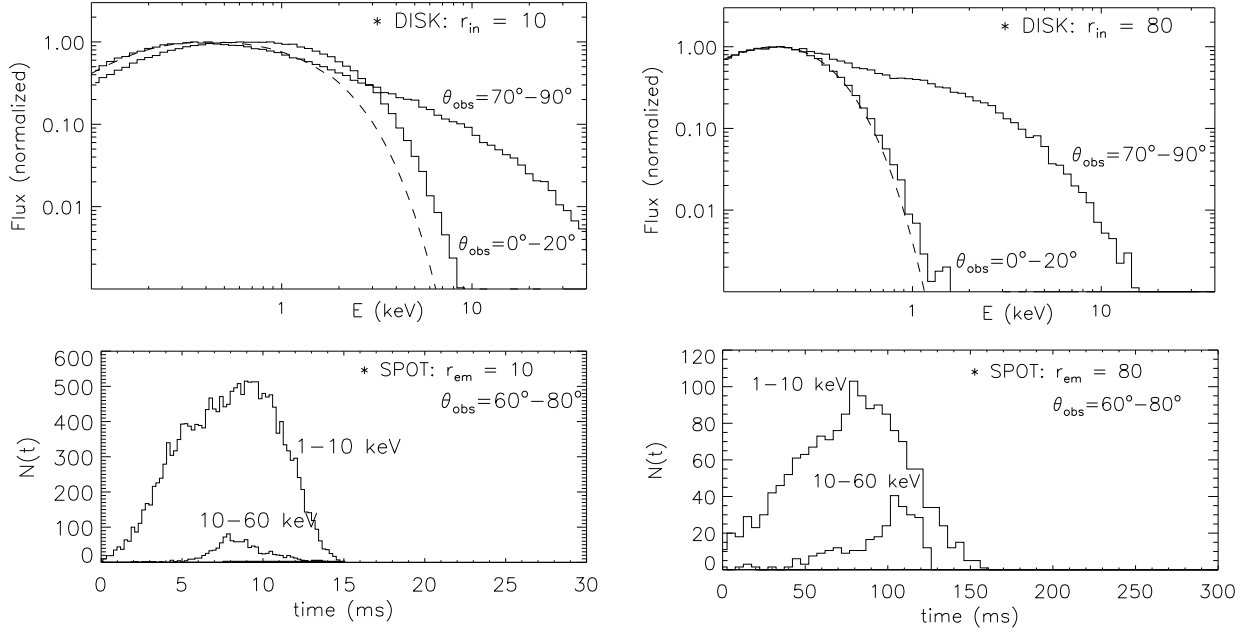}
\caption{Results of our Monte Carlo simulation.
Top panels: energy spectrum of the radiation emitted by a
standard disk (left) and by a truncated disk (right). The radiation
is partly Comptonized by the jet and forms an exponentially decreasing
hard tail whose extension depends on the angle of observation.
With the choice of the model parameters quoted in the figures, the spectral form is similar
to an ''intermediate case''. However, larger Lorentz factor and/or a higher mass loss rate
of the jet can lead to even more extended hard tails.
Bottom Panels: the figures show the differences in the arrival time of the reprocessed radiation
emitted by a localized thermal source and observed at large inclination angle.
\label{Fig4}}
\end{figure}

To compute the timing properties and the emitted spectrum respectively two separate 
approaches have been adopted.
In a first series of computations we considered a thermal point
source at $T_{em}=10$ keV,  orbiting on the equatorial plane at $r_{em}$. In this case only the scattered
radiation was considered, because it constitutes the variable component of the observed flux.
The arrival time of each photon was calculated evaluating the total path length from the point of emission to
the fiducial infinity. In a second series of runs we computed the total spectrum storing all photons
emitted by the adjacent elementary rings of the truncated disk. Most of these photons arrive directly
to infinity and form a multicolor black body spectrum, corrected for the Doppler and
gravitational effects.
Only a small, but significative fraction of the radiation emitted by the disk is scattered by the jet.
As illustrated in the top panels of Figures~\ref{Fig4}, in presence of an efficient
Comptonization by bulk motion the resulting spectrum is sensibly modified with respect to the
unperturbed multicolor distribution. In fact, despite the fact that scattered photons are a minority
of all emitted photons, they dominate the high energy part of the spectrum.
This effect is more evident when the source is far out ($r_{in}\gg 6$) and the angle of observation is
large, because it is a direct consequence of the special geometry of the configuration and of the radial
decrement of the gas density in the jet. In fact, a straightforward calculation shows that, under
these conditions, photons emitted by the disk at $\, r\,$ are mostly scattered between
$\, z_1\approx r/2\, $ and  $\, z_2\approx 4 r$,
while, at the same time,  photons gain or loose energy if $\,\thetao \,$ is larger or smaller than
$\, \arctan( r/z)$ [see eq.~(\ref{compt})].

It should be noted that, as with thermal Comptonization,
the resulting shift of the spectrum towards higher energy might cause an underestimation of the
color radius of the disk resulting from observations. Whether the color radius should or should not be
a good estimate the actual inner edge of the disk still remains an open question. Indeed,
several observations and theoretical investigations  seem to play in favor of larger values of the
inner edge of the disk \citep{ver02,mer00}. On the other hand, it is known that
the energy spectra of microquasars in their high-state may be fitted with an optically thick
accretion disk with a disk color temperature that is significantly higher then other non-jet
black hole candidates, such as Cyg X--1 and LMC X--3 \citep{ebi01}.

Finally, bottom Panels of Figure~\ref{Fig4} show the effect of the scattering on the time delay
between hard ($10 < h\nu < 60$ keV) and soft ($1< h\nu < 10 $ keV) photons. As expected, the delay
turns out to be positive when the source is located far from the axis, while it is negative for small
enough values of $r_{em}$. The amplitude of the delay is consistent with observations.

To conclude this section, we wish to emphasize that the mechanism at the basis of
our model requires the presence of collimated outflows. While radio observations of microquasars
seem to suggest a strong link between low/hard states and jets, the situation is still uncertain
concerning the high/soft case. Nevertheless, this may not be unreasonable considering that
sporadic plasma emission could occur also during high states without
forming visible radio jets, unless some conditions on density, magnetic field,
time-scales of electron acceleration, etc, are satisfied.

Clearly, in absence of powerful ejections of matter the observed spectrum has the
form of a multitemperature black body (we do not consider here the possible presence of a hot corona
above the disk). As discussed above, Comptonization by bulk motion of the radiation emitted by the
disk causes an increase of the disk color temperature. The effectiveness of this process depends
largely on the bulk Lorentz factor. However the spectral form is partly affected also by
the mass loss rate $\,\dot m$, mainly because an increase/decrease of $\,\dot m\, $
increases/decreases the extension of the optically thick portion of the jet. Therefore, in our
interpretation we expect that what we are observing
is the result either of the particular state of the accretion disk or of the physical conditions
of plasma ejections.

It is interesting to note that GRS 1915+105 is observed in three basic states A, B and C \citep{bel00}.
State A is characterized by a lower disk temperature. State B (the very high state) corresponds to an
accretion disk like the high/soft state A, but with a higher temperature and a steep power-law component.
State C (the low/hard state) corresponds to the instability period. Are these three classes associated with
configurations formed by a disk extending down the most stable orbit without any appreciable
mass ejections (state A), a full disk like state A but with a moderate collimated outflow (state B)
and, finally, a configuration formed by a disk missing of its innermost part plus a compact and
extended plasma jet (state C)? Answering this crucial question is premature at this stage because it
requires a more accurate modeling of both the disk and the jet structures, as well as a detailed
multiwavelength analysis of the evolutionary sequence of X--ray binaries. We expect to do this
in the future.

We further note that inverse Comptonizing processes accompanying discrete ejection events might
produce strong oscillations of the hard part of the spectrum on very short timescales, leaving
practically unchanged the disk soft component.
This effect might be able to account for the observed fast A-B/B-A transitions in GRS 1915+105
\citep{bel00}.

\section{Discussion and conclusions}\label{sec-discuss}

The high mass loss rates currently inferred from observations of microquasars imply that
the inner region of the relativistic outflow is optically thick to electron scattering.
If this is the case, under some conditions Comptonization of the radiation emitted by an accreting
disk produces, by bulk motion of a relativistic jet, observable
effects that might account for a number of peculiar properties of microquasars. 
If confirmed, the model proposed here
might shed light on the surprising link between the QPO phenomenon and the overall emission
properties of these sources.
It is remarkable that the different signs for the lags of different harmonics
find here a natural explanation, at variance with thermal Comptonization models which produce lags
with the same sign for all the harmonics.
Moreover, from the idealized model described in Section 2, we can expect a dependence of the
of the intensity of the processed signal on the energy, since both the energy of the scattered
photons and the (geometrical) cross section of the jet varies with the height. However, to check if
this is able to reproduce the tight power/energy correlation for the high-frequency QPO's, with
the correct sense, requires a more accurate and extensive Monte Carlo analysis. This is
matter of a future work.

Finally we may note that, if the speed of ejecta is large enough, the Comptonization of photons by
bulk motion leads to the formation of a hard tail, without the high-energy cutoff which distinguishes
the thermal models. This because there no selective effect due to the decreasing of the Klein-Nishina
cross section with the electron velocity.
It is also very interesting to note that
the need of a large inclination angle of the jet is a discriminating aspect of our model.
The discovery of systems with alternating phase lags but with no evidence of
relativistic outflows, or with jets observed at small angles, would rule out the present model,
playing in favor of a thermal interpretation.


\acknowledgments

It is a great pleasure to thank
Luca Zampieri and Silvia Zane for many useful comments and discussions. Many thanks also
to the anonymous referee for the careful reading of the manuscript and his wise comment.

\clearpage

\clearpage

\clearpage

\clearpage

\begin{thebibliography}{}
\bibitem[Belloni et al.(1997)]{bel97} Belloni, T., Mendez, M, King, A.R., van der Klis, M.,
van Paradijs, J., 1997, \apjl, 479, L145
\bibitem[Belloni et al.(2000)]{bel00} Belloni, T., Klein-Wolt, M., Mendez, M, King, A.R.,
van der Klis, M., van Paradijs, J., 2000, \aap, 355, 271
\bibitem[Chen et al.(1997)]{che97} Chen, X., Swank, J.H. \& Taam, R.E. 1997, \apjl, 477, L41
\bibitem[Ebisawa et al.(2001)]{ebi01} Ebisawa, K., Kubota, A., Mizuno, T. $\dot{\rm Z}$ycki, P.
2001, Astrophysics and Space Science 276 (suppl.), 11.
\bibitem[Fender(1999)]{fen99} Fender R.P. et al. 1999, \mnras, 304, 865
\bibitem[Gierlinski et al.(1999)]{gie99}  Gierlinski, M., Zdziarski, A.A.,
 Poutanen, J., Coppi, P., Ebisawa, K., Johnson, W.N., 1999, \mnras, 309, 496
\bibitem[Greiner et al.(1996)]{gre96} Greiner, J., Morgan, E. \& Remillard, R.A. 1996, \apjl, 473, L107
\bibitem[Greiner et al.(2001)]{gre01} Greiner, J., Cuby, J.C. \& McCaughrean, M.J., 2001, Nature,414, 522
\bibitem[Heinz and Begelman(2000)]{hei00} Heinz, S. and Begelman M.C., 2000, \apj, 535,104
\bibitem[Laurent and Titarchuk(2001)]{lau01} Laurent, P. \& Titarchuk, L., 2001, \apjl, 562, L67
\bibitem[Muno et al.(1999)]{mun99} Muno, M.P., Morgan, E., \& Remillard, R.A. 1999, \apj, 527, 321
\bibitem[Merloni et al.(2000)]{mer00} Merloni A., Fabian A.C., \& Ross R.R., 2000, \mnras, 313, 193 
\bibitem[Mirabel and Rodr\`\i guez(1994)]{mir94} Mirabel, I.F. \& Rodr\`\i guez, L.F. 1994, \nat, 371, 46
\bibitem[Morgan et al.(1997)]{mor97} Morgan, E., Remillard, R.A. \& Greiner, J. 1997, \apj, 482, 993
\bibitem[Nobili et al.(2000)]{nob00} Nobili L., Turolla, R., Zampieri, L. \& Belloni, T., 2000, \apjl, L137
\bibitem[Reig et al.(2000)]{rei00} Reig, P. Belloni, T., van der Klis, M.,
Méndez, M., Kylafis, N. D. and Ford, E. C., 2000, \apj, 541, 883
\bibitem[Reig et al.(2001)]{rei01} Reig, P., Kylafis, N.D., Spruit, H.C., 2001, \aap, 375, 155
\bibitem[Shakura and Sunyaev(1973)]{sha73}Shakura, N.J. \& Sunyaev, R.A. 1973, \aap, 24, 337
\bibitem[Sobczak et al.(2000)]{sob00} Sobczak et al. 2000, \apj, 531, 537
\bibitem[Tagger and Pellat(1999)]{tag99} Tagger, M. \& Pellat, R., 1999 \aap, 349, 1003
\bibitem[Titarchuk \& Shrader(2002)]{tit02} Titarchuk, L. \& Shrader, C.R., 2002, \apj, 567, 1057
\bibitem[Verni\`ere et al.(2002)]{ver02} Verni\`ere, P., Rodriguez, J. \& Tagger, M., 2002, \aap, to appear
\bibitem[Wijnands et al.(1999)]{wij99} Wijnands, R., Homan, J. \& van der Klis, M., 1999, \apjl, 526, L33
\bibitem[Zdziarski et al.(2001)]{zdz01} Zdzdiarski A.A., Grove, J.E., Poutanen, J., Rao, A.R. \&
Vadawale S.W., 2001, \apjl, 554, L48

\end{thebibliography}
\end{document}